\documentclass[a4paper]{jpconf}
\usepackage{graphicx}
\usepackage{amsmath}
\usepackage{url}
\usepackage{microtype}
\usepackage{color}

\begin{document}
\title{Centrality and rapidity dependence of inclusive pion and prompt photon production in p+Pb collisions at the LHC with EPS09s nPDFs}

\author{\underline{I.~Helenius$^{1,2}$}, K.~J.~Eskola$^{1,2}$ and H.~Paukkunen$^{1,2}$}

\address{$^1$Department of Physics, P.O. Box 35 (YFL), FI-40014 University of Jyv\"{a}skyl\"{a}, Finland\\
$^2$Helsinki Institute of Physics, P.O. Box 64, FI-00014 University of Helsinki, Finland}

\ead{ilkka.helenius@jyu.fi
}

\begin{abstract}
The centrality dependencies of the inclusive neutral pion and prompt photon nuclear modification factors for p+Pb collisions at the LHC are studied using a spatially dependent set of nuclear PDFs, EPS09s. The calculations are performed at mid- and forward rapidities searching for  an observable which would optimally probe the spatial dependence of the nuclear PDFs. In addition, we discuss to which $x$ values of the  nucleus the different observables are sensitive.
\end{abstract}

\section{Introduction}
In the framework of collinear factorization \cite{Collins:1989gx,Brock:1993sz}, the cross section for producing a hard parton $k$ in a nuclear collision can be calculated as
\begin{equation}
\mathrm{d} \sigma^{AB \rightarrow k + X} = \sum\limits_{i,j,X'} f_{i}^A(x_1,Q^2) \otimes f_{j}^B(x_2,Q^2) \otimes \mathrm{d}\hat{\sigma}^{ij\rightarrow k + X'} + \mathcal{O}(1/Q^2),
\label{eq:factorization}
\end{equation}
where $f_{i}^A(x,Q^2)$s are the nuclear parton distribution functions (nPDFs) describing the number distributions of parton $i$ at given values of momentum fraction $x$ and scale $Q$ for nucleus $A$. The partonic component $\mathrm{d}\hat{\sigma}^{ij\rightarrow k + X'}$ can be calculated using perturbative QCD (pQCD). The key assumption here is that, in the same way as the free proton PDFs, the non-perturbative nPDFs are process independent, so that they can be determined from a certain collection of experimental data and later on used to study different processes. It has been observed that the nPDFs, if they exist, are modified with respect to the free proton PDFs \cite{Eskola:2009uj,deFlorian:2011fp,Hirai:2007sx,Kovarik:2013sya}. This modification  can be made explicit by decomposing the nPDFs as
\begin{equation}
f_{i}^{A}(x,Q^2) = R_{i}^{A}(x,Q^2) \, f_{i}^{N}(x,Q^2),
\end{equation}
where $f_{i}^{N}(x,Q^2)$ is the free nucleon PDF and $R_{i}^{A}(x,Q^2)$ the corresponding nuclear modification. Introducing a non-perturbative input at the initial scale $Q_0$ and using the DGLAP evolution equations one can then determine $R_{i}^{A}(x,Q^2)$ via standard procedures of global analyses. Nowadays, there are several nPDF sets on the market which are reviewed e.g. in Refs.~\cite{Eskola:2012rg, Paukkunen:2014nqa}.

Each collision of two heavy nuclei takes place at a specific impact parameter $\mathbf{b}$, which is the vector in the transverse plane between the centers of the colliding nuclei. The value of the impact parameter determines the geometry (centrality) of the collision: if the impact parameter is small the amount of interacting nuclear matter is large (central collision) and if it is large, only the edges of the nuclei collide (peripheral collision). The inclusive hard particle yield in a nuclear collision is typically assumed to be proportional to the amount of interacting matter which is given by the nuclear overlap function $T_{AB}(\mathbf{b})$:
\begin{equation}
T_{AB}(\mathbf{b}) = \int \mathrm{d}^2 \mathbf{s} \,T_A(\mathbf{s+b/2}) \, T_B(\mathbf{s-b/2}),
\end{equation}
where $T_A(\mathbf{s})$ is the nuclear thickness, obtained by integrating the nuclear density profile over the beam direction.

As the thickness of the nucleus is not constant in the transverse plane, one would presume that also the nPDFs should somehow depend on the position of the nucleon inside the nucleus. However, all global fits of nPDFs have considered only minimum bias collisions (no cuts on centrality) and hence these cannot be fully consistently used for more restricted centrality classes. In Ref.~\cite{Helenius:2012wd} we addressed this problem by developing a model framework for spatially dependent nPDFs which is described in the next section. Then, in section 3, we present some predictions concerning the centrality dependence for inclusive neutral pion and prompt photon production in p+Pb collisions at the LHC at different rapidities, discuss which observable would be ideal to constrain the spatial dependence of the nPDFs, and which values of $x$ we probe at different rapidities. More predictions for p+Pb collisions at the LHC with other approaches can be found from compilation \cite{Albacete:2013ei}.

\section{Framework}

\begin{figure}[htb]
\begin{minipage}[t]{0.48\linewidth}
\centering
\includegraphics[width=\textwidth]{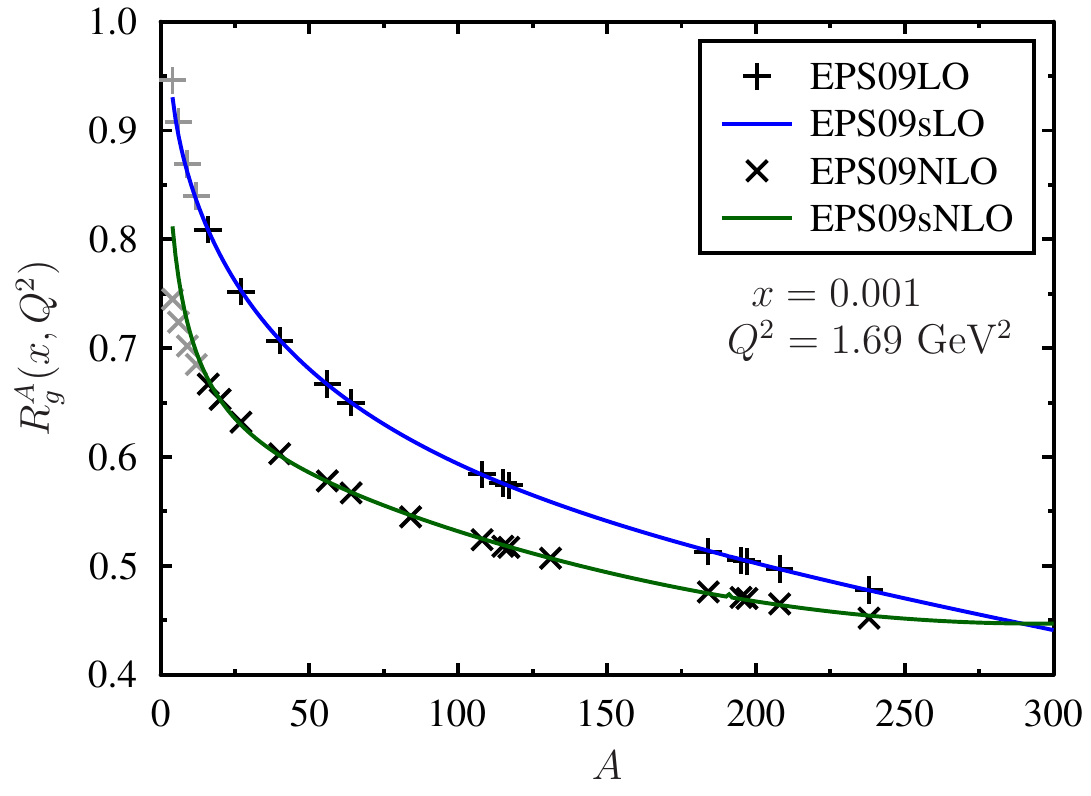}
\vspace{-15pt}
\caption{The spatially averaged (EPS09s) and the minimum bias (EPS09) nuclear modifications for gluons with fixed $x$ and $Q^2$ for leading (LO) and next-to-leading order (NLO) sets as a function of $A$. From \cite{Helenius:2012wd}.}
\label{fig:R_g_A_fits}
\end{minipage}
\hspace{0.02\linewidth}
\begin{minipage}[t]{0.48\linewidth}
\centering
\includegraphics[trim = 10pt 0pt 0pt 0pt, clip, width=\textwidth]{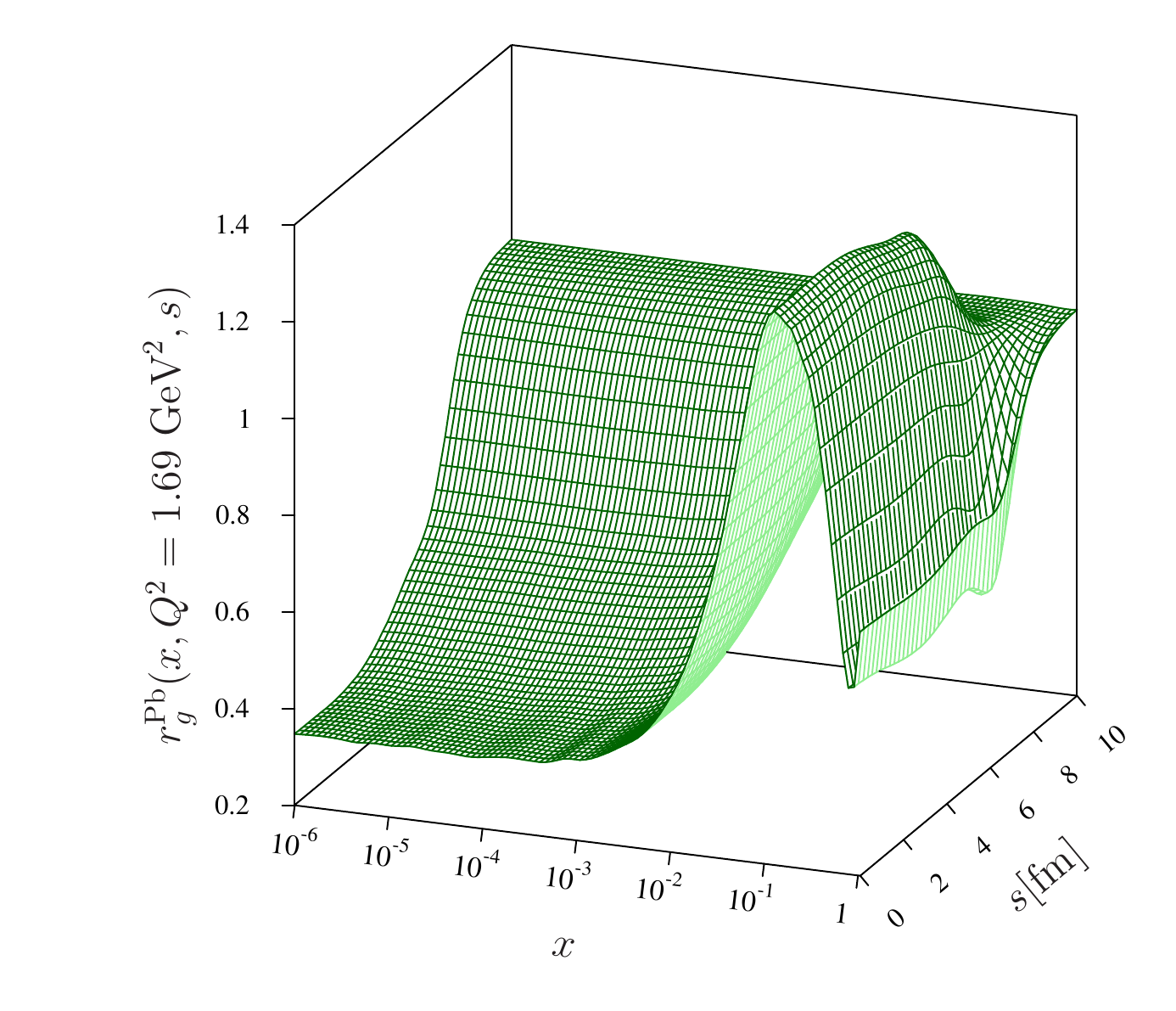}
\vspace{-15pt}
\caption{Spatially dependent NLO nuclear modification for gluons in the Pb nucleus from EPS09s as a function of $x$ and the transverse distance $s=|\mathbf{s}|$ at the initial scale $Q_0^2$. From \cite{Helenius:2012wd}.}
\label{fig:R_g_3d}
\end{minipage}
\vspace{-8pt}
\end{figure}

A natural requirement for the spatially dependent nuclear modifications $r_i^A(x,Q^2,\mathbf{s})$ is that
they should reduce to the minimum bias modification $R_{i}^{A}(x,Q^2)$ upon taking the spatial average,
\begin{equation}
R_{i}^{A}(x,Q^2) \equiv \frac{1}{A}\int \mathrm{d}^2 \mathbf{s} \,T_A(\mathbf{s})\,r_{i}^{A}(x,Q^2,\mathbf{s}).
\end{equation}
This restriction does not, however, constrain the functional form of the spatial dependence in any way, and we need to make an assumption for this. We have chosen a power series of the nuclear thickness function:
\begin{equation}
r_i^A(x,Q^2,\mathbf{s}) = 1 + \sum_{j=1}^{n} c_{j}^{i}(x,Q^2)\,[T_A(\mathbf{s})]^j.
\end{equation}
In this ansatz, all the $A$ and $\mathbf{s}$ dependencies are dictated by the thickness function and the fit parameters $c_{j}^{i}(x,Q^2)$ depend only on $x$ and $Q^2$. Thus, we can obtain the values for $c_{j}^{i}(x,Q^2)$ by using an existing set of minimum bias nuclear modifications through minimizing $\chi^2$ defined as
\begin{equation}
\chi^2_i(x,Q^2) = \sum_A \left[\frac{R^{A}_{i}(x,Q^2) - \frac{1}{A}\int \mathrm{d}^2 \mathbf{s}\, T_A(\mathbf{s})\,r^{A}_{i}(x,Q^2,\mathbf{s})}{W^{A}_{i}(x,Q^2)} \right]^2,
\end{equation}
where the sum runs over several different nuclei $A$. We have performed this procedure for the EKS98 \cite{Eskola:1998df} and EPS09 \cite{Eskola:2009uj} globally analyzed nPDF sets and we refer to the fits obtained here as EKS98s and EPS09s, respectively. For EKS98s we used the relative error as a weight ($W^{A}_{i}(x,Q^2) = 1 - R^{A}_{i}(x,Q^2)$) and for EPS09s this was set to unity. For both of these sets we were able to obtain accurate fits with four first non-trivial terms in the power series ($n=4$) in the whole kinematic region considered, see an example in Figure \ref{fig:R_g_A_fits}. The fitting was performed also for the EPS09 error sets to assess the uncertainties. An example of the resulting spatially dependent nuclear modification is plotted in Figure \ref{fig:R_g_3d} below.

\section{Results}

\subsection{Centrality classes}

We define the centrality classes in terms of impact parameter intervals which are calculated using the optical Glauber model. For the p+Pb collisions at the LHC with $\sqrt{s_{NN}}=5.0\,\rm{TeV}$ we use $\sigma_{\rm inel}^{NN} = 70\,\rm{mb}$ \cite{Antchev:2011vs} and, assuming a point-like proton ($T_{\rm pPb}=T_{\rm Pb}$), we obtain the centrality classes listed in table \ref{tab:pPb}. For p+Pb collisions with $\sqrt{s_{NN}}=8.8\,\rm{TeV}$ we have used $\sigma_{\rm inel}^{NN} = 76\,\rm{mb}$.
\begin{table}[tbh]
\caption{The impact parameter intervals and average number of binary collisions for p+Pb collisions with two collision energies for four different centrality classes.}
\begin{center}
\lineup
 \begin{tabular}{lllllll}
\br
& \multicolumn{3}{l} {$\sqrt{s_{NN}}=5.0\,\rm{TeV}$ } & \multicolumn{3}{l} {$\sqrt{s_{NN}}=8.8\,\rm{TeV}$} \\
& $b_1 \textrm{ [fm]}$ & $b_2 \textrm{ [fm]}$ & $\langle N_{\rm bin}\rangle$ & $b_1 \textrm{ [fm]}$ & $b_2 \textrm{ [fm]}$ & $\langle N_{\rm bin}\rangle$ \\
\mr
$\00-20\,\%$  & $0.0$   & $3.471$ & $14.24$ & $0.0$   & $3.494$ & $15.44$ \\
$20-40\,\%$ & $3.471$ & $4.908$ & $11.41$ & $3.494$ & $4.941$ & $12.31$ \\
$40-60\,\%$ & $4.908$ & $6.012$ & $\07.663$ & $4.941$ & $6.052$ & $\08.165$ \\
$60-80\,\%$ & $6.012$ & $6.986$ & $\03.680$ & $6.052$ & $7.029$ & $\03.810$ \\
\br
\end{tabular}
\end{center}
\label{tab:pPb}
\end{table}

\subsection{Centrality dependent nuclear modification factor}

Incorporating the spatially dependent nPDFs into the factorization formula (\ref{eq:factorization}) and multiplying this with the nuclear overlap function gives the yield of hard partons $k$ corresponding to a specific impact parameter $\mathbf{b}$ in a collision of nuclei $A$ and $B$ as follows:
\begin{equation}
\begin{split}
\mathrm{d} N^{AB \rightarrow k + X}(\mathbf{b}) =&\int \mathrm{d}^2\mathbf{s} \, T_A(\mathbf{s_1}) \,T_B(\mathbf{s_2})  \sum\limits_{i,j,X'} r_{i}^{A}(x_1,Q^2,\mathbf{s_1})f_{i}^{N}(x_1,Q^2)\\ & \otimes r_{j}^{B}(x_2,Q^2,\mathbf{s_2})f_{j}^{N}(x_2,Q^2) \otimes \mathrm{d}\hat{\sigma}^{i j \rightarrow k+X'},
\end{split}
\end{equation}
where $\mathbf{s_1} = \mathbf{s} + \mathbf{b}/2$ and $\mathbf{s_2} = \mathbf{s} - \mathbf{b}/2$. The nuclear effects can be studied by comparing the cross section in nuclear collisions to the same observable in p+p collisions. For this, we define the centrality dependent nuclear modification factor as
\begin{equation}
R_{AB}^{k}({b_1},{b_2}) \equiv \dfrac{\left\langle{\frac{\mathrm{d}^2 N_{AB}^{k}}{\mathrm{d}p_T \mathrm{d}y}}\right\rangle_{b_1,b_2}}{\frac{\langle {N_{\rm bin}} \rangle_{b_1,b_2}} {\sigma^{NN}_{\rm inel}}{\frac{\mathrm{d}^2\sigma_{\rm pp}^{k}}{\mathrm{d}p_T \mathrm{d}y}}} = \dfrac{\int_{b_1}^{b_2} \mathrm{d}^2 \mathbf{b} {\frac{\mathrm{d}^2 N_{AB}^{k}(\mathbf{b})}{\mathrm{d}p_T \mathrm{d}y}}}{ \int_{b_1}^{b_2} \mathrm{d}^2 \mathbf{b} \,{T_{AB}(\mathbf{b})}{\frac{\mathrm{d}^2\sigma_{\rm pp}^{k}}{\mathrm{d}p_T \mathrm{d}y}}}.
\end{equation}
Here, $N_{\rm bin}$ is the number of binary collisions and the average is taken over the desired impact parameter interval. Working out the averages yields the expression on the right-hand side. The minimum bias $R_{AB}$ is restored by integrating over the whole impact parameter space.

\subsection{Inclusive neutral pion production}

\begin{figure}[htb]
\begin{minipage}[t]{0.49\linewidth}
\centering
\includegraphics[width=\textwidth]{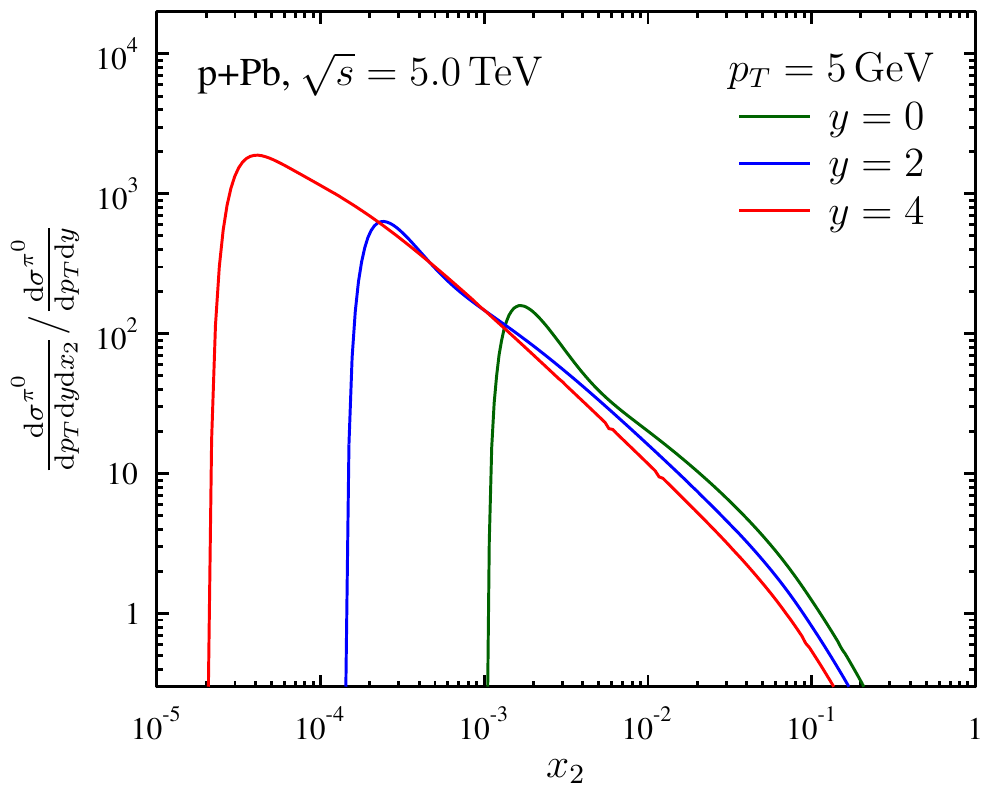}
\caption{The normalized cross section of ${\rm \pi^0}$ production in p+Pb collisions with $\sqrt{s_{NN}}=5.0\,\rm{TeV}$ for $p_T = 5\,\rm{GeV/c}$ and three rapidities, $y = 0$ (green), $y = 2$ (blue), and $y = 4$ (red) as a function of $x_2$.}
\label{fig:dsigma_x2}
\end{minipage}
\hspace{0.02\linewidth}
\begin{minipage}[t]{0.49\linewidth}
\centering
\includegraphics[width=\textwidth]{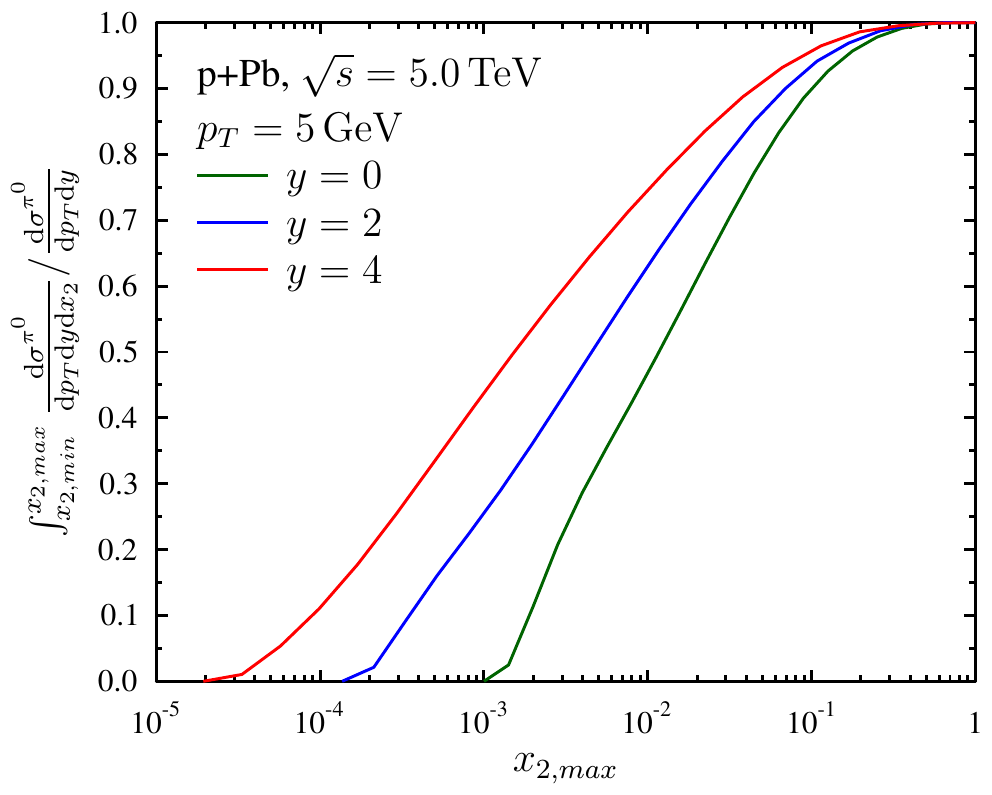}
\caption{The cumulative cross section of ${\rm \pi^0}$ production as a function of $x_2$ in p+Pb collisions with $\sqrt{s_{NN}}=5.0\,\rm{TeV}$ for $p_T = 5\,\rm{GeV/c}$ and with three rapidities, $y = 0$ (green), $y = 2$ (blue), and $y = 4$ (red).}
\label{fig:dsigma_x2_acc}
\end{minipage}
\end{figure}

The yield of single inclusive neutral pions can be calculated by convoluting the hard parton spectra with the non-perturbative parton-to-hadron fragmentation functions (FFs) $D_{{\rm \pi^0}/k}(z,Q_F^2)$
\begin{equation}
\mathrm{d} N^{\rm \pi^0}_{AB}(\mathbf{b}) = \sum_k \mathrm{d} N^{k}_{AB}(\mathbf{b}) \otimes D_{{\rm \pi^0}/k}(z,Q_F^2),
\label{eq:dNAB}
\end{equation}
where the $k$ runs over parton flavors and $z$ describes the momentum fraction carried away by the pion. As the FFs are smooth distributions in $z$, one cannot probe any specific value of $x_2$ even with fixed $p_T$ and $y$, but the cross sections get contribution from a wide range of $x_2$. This is illustrated in Figure \ref{fig:dsigma_x2} showing the normalized inclusive pion production cross section at $p_T = 5\,\rm{GeV/c}$ for three rapidities $y=0,2,4$ as a function of $x_2$. As expected, towards the forward rapidities the kinematic limit is pushed to smaller values of $x_2$, but the contribution from $x > 10^{-3}$ proves always significant. This is more clearly visible in Figure \ref{fig:dsigma_x2_acc} which illustrates how the cross section builds up when integrating over $x_2$.

\begin{figure}[htbp]
\centering
\includegraphics[trim = 0pt 2pt 0pt 10pt, clip, width=0.9\textwidth]{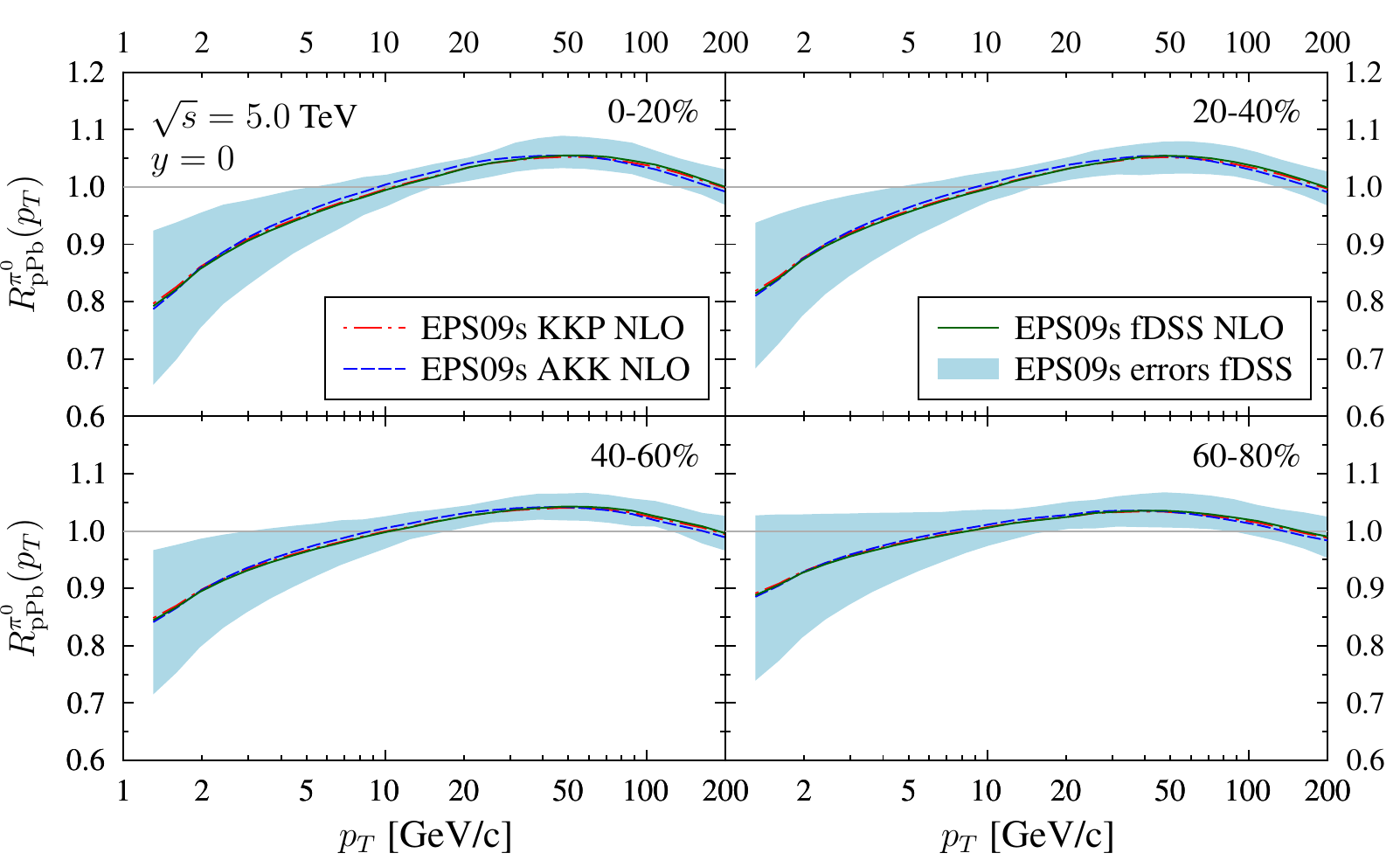}
\caption{The nuclear modification factor for $\pi^0$ production in p+Pb collisions at $\sqrt{s_{NN}}=5.0\,\rm{TeV}$ and $y=0$ in different centrality classes (different panels) with EPS09s nPDFs and KKP (red dot-dashed), AKK (blue dashed) and fDSS (green solid) FFs. The uncertainty bands are derived from the EPS09s error sets. Figure is taken from \cite{Helenius:2012wd}.}
\label{fig:pi0_R_pPb_y0}
\end{figure}

In Figure \ref{fig:pi0_R_pPb_y0}, we plot our results for the inclusive pion nuclear modification factors in p+Pb collisions with $\sqrt{s_{NN}}=5.0\,\rm{TeV}$ at midrapidity ($y=0$) for four different centrality classes using the EPS09s nuclear PDFs with different FFs: KKP \cite{Kniehl:2000fe}, AKK \cite{Albino:2008fy}, and fDSS \cite{deFlorian:2007aj}. The calculations are done with \texttt{INCNLO} \cite{Aurenche:1987fs, Aversa:1988vb, Aurenche:1999nz, incnlopage}, a public NLO code which we have modified to include also the latest FFs and to improve the convergence of the integrals at large $y$ and low $p_T$ with high $\sqrt{s_{NN}}$. All the relevant scales have been set equal to the pion $p_T$ and the uncertainty bands were calculated using the EPS09s error sets. First of all, we notice that even though there are considerable differences in the employed FFs (see e.g. \cite{d'Enterria:2013vba}), these differences tend to cancel in $R_{\rm pPb}$. With $p_T < 10\,\rm{GeV/c}$ we see some suppression as the cross sections are sensitive mostly to the small-$x$ region of the nucleus which corresponds to the shadowing in the nPDFs. At larger $p_T$ values, a small enhancement can be observed which follows from the antishadowing in EPS09s. The centrality dependence turns out as expected: in central collisions the nuclear effects are stronger than in peripheral ones. However, as the nuclear modifications are only moderate in the kinematic region considered here, also the centrality dependence is rather mild making it a challenging observable to measure.

\begin{figure}[htbp]
\centering
\includegraphics[trim = 0pt 0pt 0pt 15pt, clip, width=0.9\textwidth]{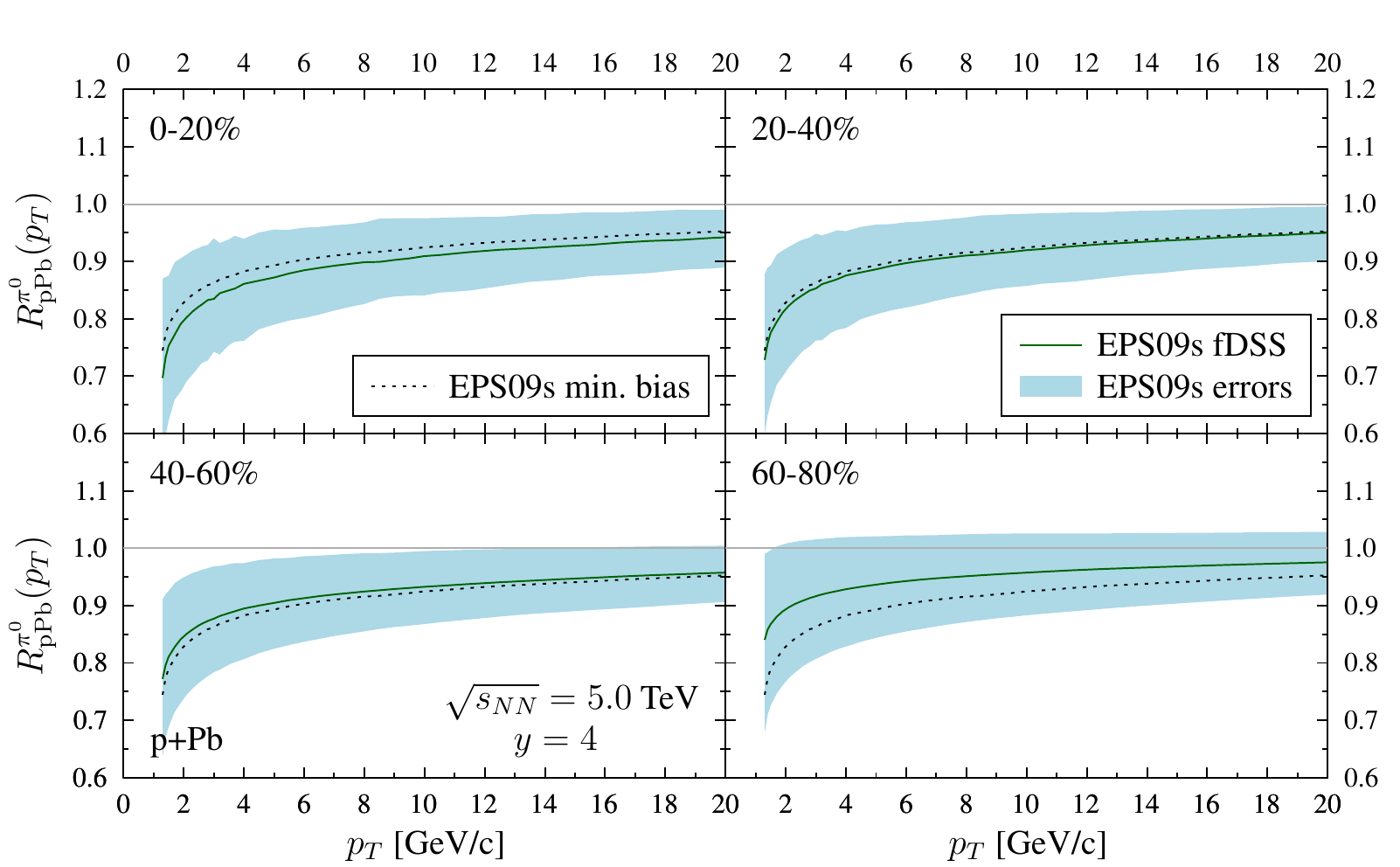}
\caption{The nuclear modification factor for $\pi^0$ production in p+Pb collisions at $\sqrt{s_{NN}}=5.0\,\rm{TeV}$ and $y=4$ for different centrality classes (different panels) with EPS09s nPDFs and fDSS FFs. The minimum bias result (dotted line) is included in each panel and the uncertainty band is derived from the EPS09s error sets.}
\label{fig:pi0_R_pPb_y4}
\end{figure}

In order to get further constraints for the nPDF centrality dependence one can consider observables in which the nuclear modifications due to the nPDFs are larger than in the mid-rapidity hadron production. One option is to consider hadron production in forward rapidities as there, according to Figure \ref{fig:dsigma_x2}, the cross sections are more sensitive to small $x_2$ values where the gluon shadowing is stronger. Therefore, Figure \ref{fig:pi0_R_pPb_y4} presents similar nuclear modification factors as in Figure \ref{fig:pi0_R_pPb_y0}, but now at $y=4$ (using the fDSS FFs only). The minimum bias result is plotted for comparison. At forward rapidity, the shadowing in the nPDFs indeed generates suppression up to higher values of $p_T$ than at mid-rapidity and the suppression is somewhat stronger. This gives rise to also a slightly more pronounced centrality dependence.

\subsection{Inclusive prompt photon production}

A better way to get sensitivity to small $x$ would be to consider final state particles that are not produced via pure fragmentation process where the convolution with the FFs largely smears the probed partonic $x$ distributions. 
Low-$p_T$ photons produced directly in the hard scattering of partons would be an ideal candidate to this end. Although it is not possible --- experimentally or strictly speaking even in the NLO calculation --- to separate these photons from those produced through similar fragmentation mechanism as pions, these two components together, which we refer to as prompt photons, still permit a more direct access to gluon PDFs than the inclusive hadrons. The prompt photon production in minimum bias p+Pb collisions was studied earlier with a similar framework in Ref.~\cite{Arleo:2011gc} and our centrality dependent results at mid-rapidity were published in  \cite{Helenius:2013bya}.

\begin{figure}[htbp]
\centering
\includegraphics[trim = 0pt 0pt 0pt 15pt, clip, width=0.9\textwidth]{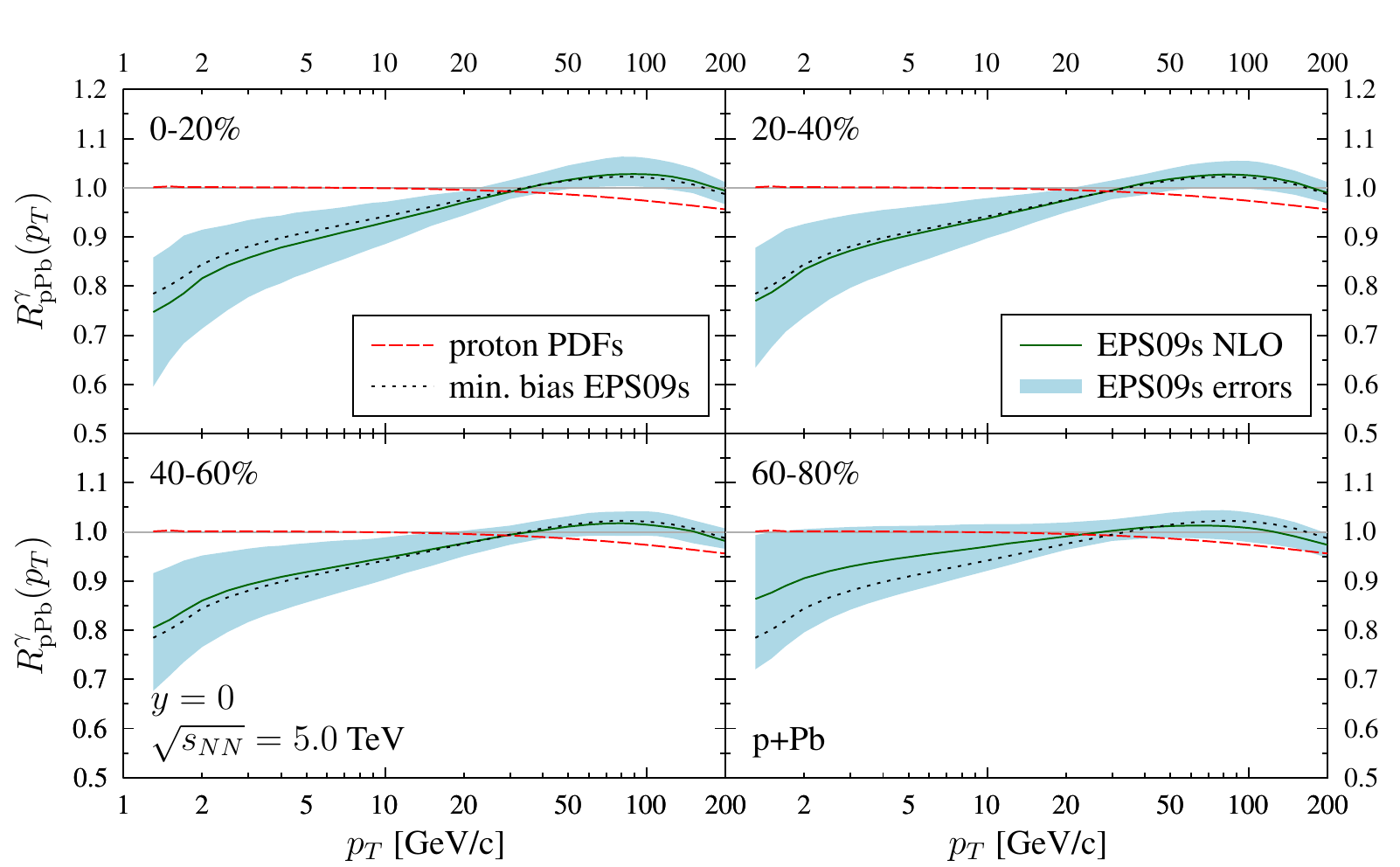}
\caption{The nuclear modification factor of inclusive prompt photon production in p+Pb collisions at $\sqrt{s_{NN}}=5.0\,\rm{TeV}$ and $y=0$ for different centrality classes (different panels) with EPS09s nPDFs. The minimum bias result is included to each panel with dotted line, the red dashed line quantifies the isospin effect and the blue uncertainty band is derived from EPS09s error sets. From \cite{Helenius:2013bya}.}
\label{fig:gamma_R_pPb_y0}
\end{figure}

To see whether an increased sensitivity to the impact parameter dependence of the nPDFs would be obtainable, the nuclear modification factors for prompt photon production in p+Pb collisions with $\sqrt{s_{NN}}=5.0\,\rm{TeV}$ at $y=0$ in different centrality classes together with the minimum bias result are plotted in Figure \ref{fig:gamma_R_pPb_y0}. For the calculation of the fragmentation component we have used the BFG II parton-to-photon FFs \cite{Bourhis:1997yu}. When comparing photons in nuclear collisions with the proton-proton collisions one should recall also the presence of an isospin effect which follows from different charge densities between the proton and neutron. This tends to yield some suppression especially at the region which is more sensitive to valence quark distributions even without any nuclear modifications in PDFs. However, from Figure \ref{fig:gamma_R_pPb_y0} we see that in the kinematic region considered here the isospin effect is negligible.
When comparing the result with the corresponding $\pi^0$ calculation (Figure \ref{fig:pi0_R_pPb_y0}) we notice that the $R_{\rm pPb}$ at $p_T < 20\,\rm{GeV/c}$ is a bit more suppressed for prompt photons, but the difference is rather small. Thus, the centrality dependence turns out to be only slightly larger than that for $\pi^0$s at mid-rapidity.

\begin{figure}[htbp]
\centering
\includegraphics[trim = 0pt 0pt 0pt 15pt, clip, width=0.9\textwidth]{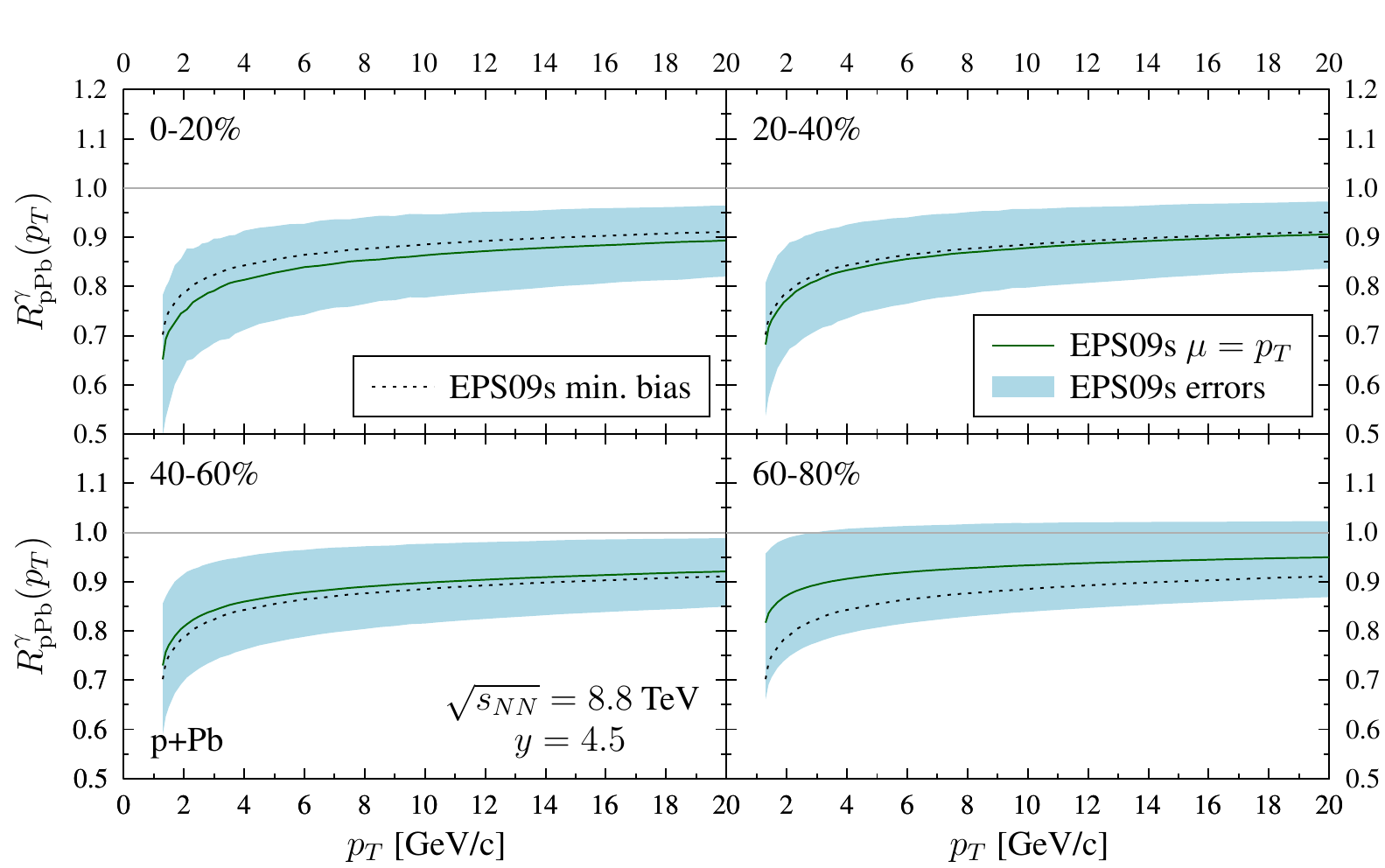}
\caption{The nuclear modification factor of inclusive prompt photon production in p+Pb collisions at $\sqrt{s_{NN}}=8.8\,\rm{TeV}$ and $y=4.5$ for different centrality classes (different panels) with EPS09s nPDFs. The minimum bias result is included to each panel with dotted line and the blue uncertainty band is derived from EPS09s error sets.}
\label{fig:gamma_R_pPb_y45}
\end{figure}

To increase the  small-$x_2$ contribution even further, we consider the prompt photon production also at forwards rapidities. In Figure \ref{fig:gamma_R_pPb_y45} we show the $R_{\rm pPb}^{\gamma}$ for p+Pb collisions with $\sqrt{s_{NN}} = 8.8\,\rm{TeV}$ at $y=4.5$ for four centrality classes and for minimum bias collisions. Even in this extreme kinematic region the maximal suppression due to the nPDFs is about 20 percent, except for the very small $p_T$ values. This follows mostly from the rapid DGLAP evolution of the NLO gluon PDFs at small $x$, and also from the presence of the fragmentation component in prompt photon production which generates contribution also from larger $x_2$ values.

\section{Summary}
Using the recently published spatially dependent nuclear PDF set EPS09s, we have calculated nuclear modification factors for p+Pb collisions at the LHC for inclusive pion and prompt photon production in different centrality classes at mid- and forward rapidities in the NLO pQCD framework. The calculations predict larger nuclear effects in central collisions compared to peripheral collisions. The centrality dependence is strongest at small $p_T$ and at forward rapidities, but still the effects are of the order $10\,\%$ at most. We notice also that even at the forward rapidities the convolution against the FFs  makes these observables less sensitive to small $x$ than could be naively expected. This could be improved by imposing an isolation cut for the prompt photons which would suppress the fragmentation component, or/and by triggering on a heavy quarks as suggested in \cite{Stavreva:2010mw}.

\ack
This work was supported by the Academy of Finland, project 133005. I. H. is supported by Magnus Ehrnrooth foundation and the PANU graduate school.

\section*{References}

\bibliographystyle{iopart-num}
\bibliography{Helenius_Grenoble}

\end{document}